\documentclass{ws-procs975x65}

\begin{document}

\title{QUANTUM PARTICLE BEHAVIOR IN CLASSICALLY SINGULAR SPACETIMES}

\author{D. A. KONKOWSKI}

\address{Department of Mathematics, U.S. Naval Academy\\
Annapolis, Maryland, 21012, USA\\
E-mail: dak@usna.edu}

\author{T.M. HELLIWELL}

\address{Department of Physics, Harvey Mudd College\\
Claremont, California, 91711, USA\\
E-mail: helliwell@HMC.edu}

\begin{abstract}
We review the classical and quantum singularity structure of a broad class of spacetimes with asymptotically power-law behavior near the origin. Quantum considerations ``heal" a large class of scalar curvature singularities.
\end{abstract}

\bodymatter

\section{Introduction}
The question addressed in this review is: What happens if instead of classical particle paths (time-like and null geodesics) one uses quantum mechanical particles to identify singularities? The answer for any asymptotically power-law space-time is given. This conference proceeding is based on articles by the authors \cite{HK} and by K. Lake \cite{Lake}.

\section{Types of Singularities}

\subsection{Classical Singularities}
A classical singularity is indicated by incomplete geodesics or incomplete paths of bounded acceleration \cite{HE, Geroch} in a maximal spacetime. Since, by definition, a spacetime is smooth, all irregular points (singularities) have been excised; a singular point is a boundary point of the spacetime. There are three different types of singularity \cite{ES}: quasi-regular, non-scalar curvature and scalar curvature. Whereas quasi-regular singularities are topological, curvature singularities are indicated by diverging components of the Riemann tensor when it is evaluated in a parallel-propagated orthonormal frame carried along a causal curve ending at the singularity.

\subsection{Quantum Singularities}
A spacetime is QM (quantum-mechanically) nonsingular if the evolution of a test scalar wave packet, representing the quantum particle, is uniquely determined by the initial wave packet, manifold and metric, without having to put boundary conditions at the singularity\cite{HM}. Technically, a static ST (spacetime) is QM-singular if the spatial portion of the Klein-Gordon operator is not essentially self-adjoint on $C_{0}^{\infty}(\Sigma)$ in $L^2(\Sigma)$ where $\Sigma$ is a spatial slice.

\section{Asymptotically Power-Law Spacetimes}
We consider a class of spacetimes that can be written in power-law metric form in the limit of small $r$, 

\begin{equation}
ds^2= -r^\alpha dt^2 + r^\beta dr^2 + C^{-2} r^\gamma d\theta^2 + r^\delta (dz + A d\theta)^2
\end{equation}

\noindent where $\beta, \gamma, \delta, C, A$ are constant parameters and the variables have the usual ranges. We are particularly interested in the metrics at small $r$, because we suppose that if the spacetime has a classical curvature singularity (and nearly all of these do), it occurs at $r=0$. \footnote{If $\alpha = \beta = \gamma = \delta = 0$, $C \neq 1$ indicates a quasi-regular singularity (a disclination) and $A \neq 0$ indicates a quasi-regular singularity (a dislocation) (see, e.g., Konkowski and Helliwell\cite{KH}).}

We can eliminate $\alpha$ by rescaling $r$ which results in two separate metric types:

\begin{itemlist}
  \item Type I:
    
\begin{equation}
ds^2 = r^\beta (-dt^2 + dr^2) + C^{-2} r^\gamma d\theta^2 + r^\delta (dz + A d\theta)^2
\ \ \ \ \ \ \ \ \ \ \alpha \neq \beta + 2.
\end{equation}
 
  \item Type II:

\begin{equation}
ds^2 = -r^{\beta + 2} dt^2 + r^\beta dr^2 + C^{-2} r^\gamma d\theta^2 + r^\delta (dz + A d\theta)^2
\ \ \ \ \ \ \alpha = \beta + 2.
\end{equation}

\end{itemlist}

\section{Classical Singularity Analysis}
Except for isolated values of $\beta, \gamma, \delta, C, A$ all of these power-law spacetimes have diverging scalar polynomial invariants if and only if $\beta > -2$. 

\subsection{Type I Spacetimes}
Lake \cite{Lake} has shown that in Type I STs $r=0$ is timelike, naked and at a finite affine distance if and only if $\beta > -1$,  implying that there is a classical singularity at $r=0$ if and only if $\beta> -1$.
 
\subsection{Type II Spacetimes} 
Likewise, Lake \cite{Lake} has shown that in Type II STs $r=0$ is null, naked and at a finite affine distance and thus is a classical singularity for all $\beta > -2$.

\section{Quantum Singularity Analysis}
To study the quantum particle propagation in these spacetimes (for simplicity, we take $A = 0$), we use massive scalar particles described by the Klein-Gordon equation and the "limit point - limit circle" criterion of Weyl \cite{RS, Weyl}. This means that, in particular, we study the radial equation in a one-dimensional Schr\"odinger form with a 'potential' and determine the number of solutions that are square integrable. If we obtain a unique solution, without placing boundary conditions at the location of the classical singularity, we can then say that the Klein-Gordon operator is essentially self-adjoint and the spacetime is QM-nonsingular.

\subsection{Type I Spacetimes}
There is a quantum singularity "bowl" in parameter space for these metrics. The bowl is bounded by (1) a bottom which is formed from a $\beta=-2$ base plane and (2) the sides which are composed of (a) two vertical planes with $\gamma+\delta=6$ and $\gamma+\delta=-2$ and (b) two tilted planes with $\delta=\beta+2$ and $\gamma=\beta+2$. Points within the bowl are QM singular; points outside the bowl are QM non-singular.

\subsection{Type II Spacetimes}
Type II STs are globally hyperbolic; the wave operator in this case must be essentially self-adjoint, so these spacetimes contain no quantum singularities. It is easy to verify this conclusion directly by checking the essential self-adjointness of the wave operator using the ``limit point - limit circle" technique.

\section{Conclusions}
A large class of classically singular asymptotically power-law spacetimes has been shown to be quantum mechanically non-singular. Invoking an energy condition (e.g., weak or strong) can eliminate more singular spacetimes, but no choice completely eradicates them.

\section{Acknowledgments}
One of us (DAK) thanks Queen Mary, University of London, where some of this work was carried out.

\bibliographystyle{ws-procs975x65}
\bibliography{ws-pro-sample}

\end{document}